# DISCRETE THERMODYNAMICS OF CHEMICAL EQUILIBRIA

B. Zilbergleyt[1]

> *"Nature creates not* genera *and* species*, but* individua*, and our shortsightedness has to seek out similarities so as to be able to retain in mind many things at the same time."*
> G.C. Lichtenberg. "The Waste Books", Notebook A, 1765/1770.

**Introduction**

These words, though ironically, express the Ockham's Razor essence [1] – to employ as less entities as possible to solve maximum of problems. Both reflect the efforts of this work to unite on the same ground some things in chemical thermodynamics that seem at a glance very unlikely.

A difference between theoretically comfortable and easy to use isolated chemical systems and real open systems[2] was recognized long ago. A notion of the open chemical systems non-ideality, offered by G. Lewis in 1907 [2], had responded to that concern. Fugacities and thermodynamic activities, explicitly accounting for this notion, were introduced for non-ideal gases and liquid solutions to replace mole fractions in thermodynamic formulas. Leading to the same habitual linear dependence of thermodynamic potential vs. activity logarithm (instead of concentration), that allowed the thermodynamic functions to keep the same appearance, passing open and closed systems for isolated entities. With this support classical theory has survived unchanged for a century. J. Gibbs is credited for a presentation of the phases in the multiphase equilibrium as a set of open subsystems with equal chemical potentials of common components [3]. Gibbs' insight hasn't impacted essentially thermodynamic tools for chemical equilibria.

Energy/matter exchange between an open system and its environment is accompanied by energy dissipation that, under a strong external impact, can be compensated only by changing the system state and even its organization. That became clear with discovery of self-organizing open chemical systems, forming dissipative structures [4]. Explanation of this phenomenon was well beyond the grasp of classical thermodynamics, and a new reality has arisen. The term "far-from-equilibrium" was coined to designate loosely a place of honorable exile from equilibrium thermodynamics for a great deal of real systems. Contemporary chemical thermodynamics literally became torn apart applying different concepts to traditional isolated systems with true thermodynamic equilibrium (TdE) as the only achievable state [5], and to the open systems with self-organization. None of the recognized models allows for transition from one to another without sacrificing model's integrity. At this point classical chemical thermodynamics, seemingly powerful and definitely elegant in applications to traditional isolated systems, looses its power and turns into a clumsy schizothermodynamics ("split thermodynamics", also in a possible idiomatic sense). In classical theory, isolated system is a heavily reigning routine, sometimes releasing the vapor with the help of closed systems, while open system actually never arrives. Treatment of the self-organizing systems employs mostly the "entropic" approach; isolated entities are usually treated within the classical "energetic" concept. Works of I. Prigogine (e.g., [4]) and his school regard the entropy production as the major factor to define the direction and the outcome of chemical processes. Such a *modus operandi* implicitly puts entropy in charge of the system behavior, in a sense replacing other thermodynamic functions. Though some authors consider the "entropic" approach more fundamental than the "energetic" [6], it works relatively well in case of not robust reactions but isn't quite capable to cover full range of chemical transformations. Discrete Thermodynamics of Chemical Equilibria (DTD) offers an effective solution to the problem, unifying both thermodynamic aspects in a unique theory on the energetic basis. DTD is the subject of this paper.

---

[1]   System Dynamics Research Foundation, Chicago, USA, e-mail: sdrf@ameritech.net.

[2]   At this point we apply general name open system to all non-isolated entities including closed systems.

**Definitions**

An isolated chemical system achieves the state of chemical equilibrium at zero rate of chemical reaction. This state coincides with TdE where reaction characteristic functions take on minimal values. *TdE is essentially a reaction state.* If system is open, additional condition for the system equilibrium with its environment includes zero uncompensated energy and material exchanges between them. *This is the state of a chemical system that harbors the chemical reaction.*

Majority of chemical reactions run at p,T=const, and standard change of Gibbs' free energy $\Delta G^0$ is a usual criterion of the reaction robustness – its ability to run up to equilibrium and achieve corresponding reaction extent. We employ instead a more informative value of the *thermodynamic equivalent of chemical transformation* $\eta$, a ratio between any participant amount, chemically transformed within the isolated system *ab initio* to TdE (with asterisk), per its stoichiometric unit

(1) $\qquad\qquad\qquad\qquad\qquad\qquad\qquad\qquad\qquad\qquad\qquad \eta_j = \Delta^* n_{kj}/\nu_{kj}.$

Given thermodynamic parameters, stoichiometric equation (and therefore the reaction thermodynamic characteristics), and the system chemical composition, $\eta_j$ is the system invariant having the same value for all reaction participants. This value arrives as the only result in thermodynamic simulation of TdE, and this is the way to find it numerically. In DTD it serves as basic reference measure for the chemical system states.

Reaction coordinate $\xi_D$ was introduced by De Donder [7]

(2) $\qquad\qquad\qquad\qquad\qquad\qquad\qquad\qquad\qquad\qquad\qquad d\xi_D = dn_{kj}/\nu_{kj}$

with dimension of mole. We re-define it as

(3) $\qquad\qquad\qquad\qquad\qquad\qquad\qquad\qquad\qquad\qquad\qquad d\xi_Z = (dn_{kj}/\nu_{kj})/\eta_j,$

or

(4) $\qquad\qquad\qquad\qquad\qquad\qquad\qquad\qquad\qquad\qquad\qquad d\xi_Z = dn_{kj}/\Delta^* n_{kj}.$

In finite differences the reaction extent $\Delta\xi_Z$ is

(5) $\qquad\qquad\qquad\qquad\qquad\qquad\qquad\qquad\qquad\qquad\qquad \Delta\xi_Z = \Delta n_{kj}/\Delta^* n_{kj},$

$\Delta n_{kj}$ is the amount of k-moles transformed in the reaction from its initial to current states. *So defined reaction extent is a dimensionless marker of equilibrium with the initial value $\Delta\xi_Z=0$ and $\Delta\xi_Z=1$ at TdE.* If $\Delta n_{kj}$ represents the k-moles amount transformed in j-reaction to its chemical equilibrium (which may not match TdE, see below), $\Delta\xi_Z$ *measures chemical equilibrium in parts of TdE*. A resembling restriction $0\leq\xi\leq1$ was discussed long ago [8] for the coordinate defined by (2) in a different context; $\xi_D$ takes on unity only if the reactants, initially taken in stoichiometric ratio (logistic end of reaction), are exhausted, and within this interval $\xi_D$ tallies the distance between the current state and the logistic end point.

We define the system shift from TdE as

(6) $\qquad\qquad\qquad\qquad\qquad\qquad\qquad\qquad\qquad\qquad\qquad \delta\xi_Z = 1-\Delta\xi_Z.$

Its sign is positive if reaction didn't reach TdE (or is moved towards the initial state) and negative if its state is beyond TdE. Obviously, $\delta\xi_Z=1$ at the initial state, $\delta\xi_Z=0$ at TdE, and in general $1\geq\delta\xi_Z\geq0$. Reaction extent, based on De Donder's coordinate, may be designed in similarity with (5); it would be linked to the newly defined value as

(7) $\qquad\qquad\qquad\qquad\qquad\qquad\qquad\qquad\qquad\qquad\qquad \Delta\xi_D = \eta_j\Delta\xi_Z.$

Further on we will use exclusively z-subscripted values omitting the subscript and retaining in writing only $\Delta_j$ and $\delta_j$. Those quantities provide for a great convenience in equilibrium analysis: DTD employs $\eta_j$ as the theory parameter (the first of two) and $\delta_j$ as its basic variable.

As it was started earlier in this paper, the DTD draws a distinctive line between the chemical reaction parameters and values and the same but relevant to the chemical system, sheltering that reaction. Typical reaction parameters are $\nu_{kj}$, $\xi_Z$ and $\Delta G^0$. Besides the above mentioned $\eta_j$, Onsager coefficients $a_{ij}$ with $i\neq j$ and the shift $\delta_j$ are examples of the system parameters. In some cases distinction between them and the hosted reaction parameters is contextual. In the following text we will mention the values relevance as needed.

*A concept, demanding the system to be divisible by parts in a way that those parts and their relations constitute the system itself is one of the pillars of the theory.* We suppose that the system equilibrium is achieved when all its subsystems are at equilibrium to each other; in other words, when each subsystem is at binary equilibrium with its compliment to the whole system. At this state each subsystem rests at its own open equilibrium. Interactions between subsystems are namely the factors leading directly to the famous Aristotle's observation that "The whole is more than the sum of its parts". Obviously, our picture of the whole chemical system equilibrium state differs from traditional notion of *detailed equilibrium* [8].

In general, subsystems and open/closed systems are synonymous. While classical theory is busy to describe exclusively the systems, we have to look closely also at their relations with the environment, that are essentially different for both types of the systems. State of the source of electrical potential, applied to an electrochemical cell, doesn't depend on the cell; the source of pumping force, applied to a laser, is totally independent on the laser, and the state of a gas torch, heating water in a closed vessel, is totally autonomous. There is no material exchange or chemical interactions between the subsystems and the energy sources in these examples; the external impact is mostly of a physical nature, and state of its source is totally independent on the state of the system experiencing the action. All those entities, usually surrounded by not permeable for matter and semi-permeable for energy boundary, fall into classical category of closed systems. Open chemical system, exchanging matter with its environment, may have no boundary at all and both may be even not separated spatially like gaseous components of different reactions in a complex chemical system. Open system is definitely a subsystem of a bigger system, both are experiencing extensive mutual impact and their states are changing until the entire system achieves equilibrium. Further on we will deal mainly with the closed systems, whose sources of external impact are abstract and represented only by the action they produce. When the environment and material exchange are considered explicitly, they both turn into mutually dependent sources of impact.

Interactions between thermodynamic systems may be formalized in *thermodynamic forces*, understood as the reasons for changes in the systems, against which they act [9]. Termodynamic force was introduced to chemical thermodynamics by De Donder as thermodynamic affinity [7], a moving power of chemical transformations

(8) $\qquad A_j = - (\delta \Phi_j / \delta \xi_j)_{x,y}$,

$\Phi_j$ stands for major characteristic functions or enthalpy, appropriate thermodynamic parameters are $x$ and $y$. Affinity definition (8) matches general definition of force in physics as a negative derivative of potential by coordinate. De Donder's affinity represents the intrinsic force, the *eugenaffinity,* moving the chemical reaction to TdE, and is a typical reaction value. Switch from $\xi_D$ to $\xi_Z$ adds only a factor $(1/\eta_j)$ to the right hand side of (8), but makes the coordinate dimensionless and the affinity dimension becomes the same as of the corresponding characteristic function.

To illustrate major ideas throughout this paper, we will mostly use a gaseous reaction

(9) $\qquad PCl_3 + Cl_2 = PCl_5$

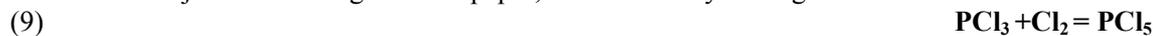

due to its simplicity and large equilibrium composition changes in a narrow temperature range.

**The Le Chatelier Response**

Open system interactions with its environment may be better formalized via relationship between the external impact and the system response. We will do that formalization based on Le Chatelier principle. Stating, that stressed system decreases thermodynamic mismatch with its environment by dodging to a state with the minimal stress allowed by external constraints, the principle describes perhaps the simplest known self-organization in nature. H. Le Chatelier (see [10]), R. Etienne [11], T. De Donder [12] tried to move that verbal principle towards the use of numbers, offering a set of moderation theorems. Unfortunately, their efforts ended up with inequalities and didn't change much the principle qualitative character. It is quite obvious that the shift from TdE must be proportional to the stress, caused by external TdF. Let's define the *Le Chatelier Response* (LCR) $\rho_j$ as power series of the system shift from TdE, proportional to TdF

(10) $$\rho_j = \Sigma_{0 \to \pi} \omega_p \delta_j^p = (1/\alpha_j) F_{je}.$$

$(1/\alpha_j)$ is a proportionality coefficient, $F_{je}$ is the symbol for TdF. The weights $w_p$ are supposed to be within the interval [0,1]; they are unknown *a priori* and we have no other choice as to equalize them to unity (besides $w_0$, see below). Expression (10) accounts for non-linearity of the system response and leads to incomparably less bulky problem solution than based on power series of the external force [8]. Variety of the system response modes is caused by the system intricacy, and $\pi$ may be identified with the *system complexity parameter* (the second parameter of two in the theory). Dimension of thermodynamic force is energy, $\delta_j$ is dimensionless, and dimension of $\alpha_j$ also should be energy. Fortunately, further derivation and its results are very similar for the LCR by (10) and its simplified linear form $\rho_j=\delta_j$ we get at $\omega_0=0$, $\pi=1$ and $\omega_1=1$, that is

(11) $$\delta_j = (1/\alpha_j) F_{je}.$$

That allows us to conduct major derivations and illustrations using (11), more complicated cases will be introduced later.

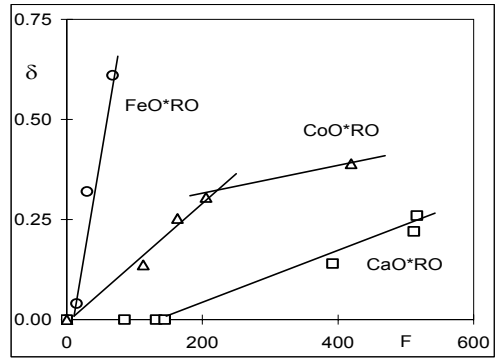

Fig.1. System shift vs. external force $(-\Delta G_f^0/\Delta_j)$, kJ/mol, 298.15K, (MeO·RO+S) series, simulation (HSC).

Hypothesis (11) occurred to be quite provable. Imagine reaction of a metal oxide MeO with a reductant Re in isolated system; at TdE $\Delta_j=1$. Suppose a double oxide MeO·RO, where only MeO still can, but RO cannot react with Re and just restricts reacting activity of the first. Let MeO·RO now react with Re. Complex chemical system MeO-RO-Re comprises two open to each other subsystems with potential reactions MeO+Re and MeO+RO, the subsystems compete for possession of MeO. Now the MeO+Re reaction outcome will be different than in the case of free oxide: acting against the subsystem MeO+Re TdF, originated from binding MeO into MeO·RO, causes $\Delta_j<1$ and $\delta_j>0$ at complex chemical equilibrium. According to (11), this shift must be proportional to the TdF.

In our system TdF=$\Delta G^0_{MeO \cdot RO}/\Delta_j$, a ratio between Gibbs' free energy standard change of the MeO·RO formation from the oxides and the (MeO+Re) reaction extent at complex (MeO·RO+Re) equilibrium. The shifts dependency on the TdF for the same MeO and various RO should be linear. The data in Fig.1 were obtained by traditional thermodynamic simulations of the oxide interactions with sulfur, conducted in two series – single oxides (MeO+S), MeO=FeO,CoO,CaO, to find $\eta_{MeO}$, and double oxides (MeO·RO+S), RO from a group SiO2,Fe2O3,TiO2,WO3, Cr2O3, to find $\eta_{MeO \cdot RO}$. TdE states in the (MeO+S) systems naturally coincide with the reference scale zero point at $\delta_j=0$; the shifts in the (MeO·RO+S) systems were calculated as

(12) $$\delta_j=1-\eta_{MeO \cdot RO}/\eta_{MeO}.$$

The shift-force linearity is a good first approximation for the initial part of the graph, in reality it is a part of a non-linear diagram as it will be shown further.

**Premises of Discrete Thermodynamics**

Presented in this paper discrete thermodynamics of chemical equilibria is based on the Le Chatelier principle in form of the LCR, De Donder thermodynamic affinity and Onsager constitutive linear equations.
Thermodynamic affinity (8) in discrete form may be put down as

(13) $$A_j = -\Delta\Phi_j/\Delta\xi_j.$$

Onsager linear constitutive equations of non-equilibrium thermodynamics, that link thermodynamic forces (TdF), or affinities in case of chemical systems − internal $A_{ji}$, acting within j-system, and external $A_{je}$, acting against it from outside − to the reaction rate [8], are

(14) $$v_j = a_{ji}A_{ji} + \Sigma a_{je}A_{je}.$$

We group all interactions within complex chemical system into dichotomial couples, each comprising a subsystem and its environment that includes in turn all other subsystems and represents the subsystem compliment to the big system. Such an approach reduces the amount of interactions to be accounted for down to the number of subsystems, adds clarity to our concept of complex systems, and makes derivations simpler. That turns (14) to

(15) $$v_j = a_{ji}A_{ji} + a_{je}A_{je},$$

$a_{je}A_{je}$ is a contribution from the j-subsystem complement [13]. At chemical equilibrium $v_j=0$

(16) $$A_{ij} + o_j A_{je} = 0,$$

The first term is the *bound affinity* [14], a non-zero remainder of the eugenaffinity, balancing the shifting force at open equilibrium and equal to it; dimensionless ratio $o_j = a_{je}/a_{ji}$ is a reduced "complimentary" Onsager coefficient. Equations (15) and others, derived from it, mean a balance between the internal and the external thermodynamic forces that holds the j-system in this state. Examples of the external TdF with regards to subsystem-environment relations were given earlier; the double-oxide task represents another instance of the open system and variable TdF.
The chemical system deviates from TdE until the bound affinity balances the TdF, i.e. $\alpha_j\rho_j = o_j A_{je}$. Substituting $o_j A_{ji} = -(\Delta\Phi_j/\Delta_j)_{x,y}$ and placing this new expression into (10), after multiplying both sides by $\Delta_j$ we obtain a new condition of chemical equilibrium

(17) $$\Delta\Phi_j(\eta_j,\delta_j)_{x,y} + \alpha_j\rho_j\Delta_j = 0,$$

In the linear case (17) turns to

(18) $$\Delta\Phi_j(\eta_j,\delta_j)_{x,y} + \alpha_j\delta_j(1-\delta_j) = 0.$$

Expression (17), the basic expression of the discrete thermodynamics, is a sort of logistic map [6] that, unlike the traditional versions, employs the system shift from TdE as variable.

**The Map of States of the Chemical System at p, T=const**

Most chemical processes run at p,T=const, the characteristic function is Gibbs' free energy, and (18) turns to

(19) $$\Delta G_j(\eta_j,\delta_j) + \alpha_j\delta_j(1-\delta_j) = 0,$$

leading to intermediate expression

(20) $$[\Delta G_j^0 + RT\ln\Pi_j(\eta_j,\delta_j)] + \alpha_j\delta_j(1-\delta_j) = 0.$$

Now we have a general map for isothermobaric chemical equilibrium. Because the $\alpha_j$ dimension is energy, one may interpret it as $\alpha_j = RT_{alt}$ with some sort-of-temperature value, an *alternative* temperature with dimension in Kelvins that provides for the map uniformity. Also, $\Delta G_j^0 = -RT\ln K$, or $\Delta G_j^0 = -RT\ln\Pi_j(\eta_j,0)$. Being divided by RT, map (20) turns to

(21) $$\ln[\Pi_j(\eta_j,0)/\Pi_j(\eta_j,\delta_j)] - \tau_j\delta_j(1-\delta_j) = 0,$$

where $\tau_j = T_{alt}/T$, the isothermobaric "growth" factor, defining the increase of the system deviation from TdE. Like in the Verhulst model of bio-populations [15], its numerator $RT_{alt}$ represents external impact on the system ("demand for prey" [16]) while the denominator $RT$ is a measure of the system resistance to that impact. Increase of $\tau_j$ drives chemical system towards complex behavior with bifurcations and eventual chaos [17]. That's why in some previous publications we also called this factor *reduced chaotic temperature*. Ratio $\Pi_j(\eta_{kj},0)/\Pi_j(\eta_{kj},\delta_j)$ is a reverse value of the relative size of "chemical population" – a ratio of molar parts product in the isolated system at TdE to the same function for the equilibrium open system under external impact. The numerator, $\Delta G_j^0/RT$, corresponds to maximum undisturbed population size, a *carrying capacity of the chemical system,* similar to the carrying capacity of the areal in bio-populations.

Map (21) specifies conditions of chemical equilibrium *in any isothermobaric chemical system,* both open and isolated as well. Its first term is the classical change of the j-system Gibbs' free energy, a value, relevant to j-reaction[1]. The second, parabolic term reflects interactions between the j-subsystem and its environment. This term makes difference between the classical and our theory and causes a rich variety of "far-from-equilibrium" behaviors. *Together they represent nothing else but full change of Gibbs' free energy in open chemical system, equal to zero when the system is at equilibrium with its environment. At $\delta_j=0$ the map turns into classical change of Gibbs' free energy for isolated state*, and defines TdE. We will use the term *map of states* for expression (17) and its particular forms; the reader will see that it holds more sense than just a substitution for the "equation of state".

**Equilibrium Constant as Universal Parameter of Chemical Equilibria**

To generalize previous results and for the future use, let's introduce a shift function
$$(22) \qquad \varphi(\delta_j,\pi) = \rho_j\Delta_j = \rho_j(1-\delta_j).$$
Expression (17) turns to
$$(23) \qquad \Delta\Phi_j(\eta_j,\delta_j)_{x,y} + \alpha_j\varphi(\delta_j,\pi) = 0,$$
and (21) may be rewritten as
$$(24) \qquad \Pi_j(\eta_j,0) = \Pi_j(\eta_j,\delta_j)\exp[\tau_j\varphi(\delta_j,\pi)].$$
Now one can clearly see operational meaning of the map of states – it maps population of the isolated j-system at TdE, $\delta_j=0$, into the same system population at open equilibrium with $\delta_j \neq 0$: *states of the open chemical system can be deduced from its isolated state*. Obviously, the left hand side of (24) is the j-reaction constant of equilibrium $K_j$ by definition. We arrive at a simple general rule for equilibrium states in isothermobaric chemical systems
$$(25) \qquad K_j = \Pi_j(\eta_j,\delta_j)\exp[\tau_j\varphi(\delta_j,\pi)]$$
including TdE at $\varphi(\delta_j,\pi)=0$. Analytical or tabulated relations between $\delta_j$ and $\tau_j$ allow us to calculate the chemical system composition at any $\delta_j$ given equilibrium constant. *Equilibrium constant is a quantitative characteristic of the chemical system carrying capacity regarding its chemical populations.*

Traditionally expressions (21) and (25) are supposed to map the n-subscribed values into (n+1); in our case both sides are subscribed identically. The reason is simple – our whole theory is built up around chemical equilibrium, and iterative calculations end up at $\delta_{n+1}-\delta_n < \varepsilon$, i.e. at the point where both values cannot be distinguished at given measure of accuracy. This is how the real system approaches its equilibrium with a measure $\varepsilon$ of the system oscillations around it. As it follows from the derivation and structure of map (21), the open system behavior and states are governed by the resultant of the set of thermodynamic forces acting against it. Internal equilibrium, i.e. TdE in a subsystem makes sense only as a reference state which the subsystem takes on being turned into an isolated entity. As a result, the notion of detailed equilibrium, a hypothesis that traditionally plays key role in some approaches to complex chemical equilibrium, should be replaced by the notion of the system equilibrium as a set of interrelated local binary equilibria, each defined for a certain subsystem with its own map of states (17). At the same time, it follows neither from this map nor from its derivates that the local equilibrium of an arbitrary subsystem demands the whole system to be at equilibrium. If this is correct, it widens potential applications of the developed theory. Equilibria in open systems are pretty much the stationary states. It should be kept in mind that stationary states of chemical systems may be of two types – equilibrium and non-equilibrium [8], and by virtue of its derivation with nullified material flow the map (17) and its particular versions relate to *equilibrium stationary states only*. Though these states are thermodynamic in essence due to zero balance of thermodynamic forces, acting against the system, we reserve traditional term "true thermodynamic equilibrium", or TdE for isolated systems with only one chemical transformation within, using more general term "chemical equilibria" for open chemical systems and their environment. Non-equilibrium stationary states represent a separate task that may be solved by discrete thermodynamics.

---

[1] As it will be explained later, map (21) is valid for the strong type of chemical systems.

**The Map of States of the Chemical System at p, S=const**

Another important group of chemical transformations are adiabatic processes occurring at p, S=const, their characteristic function is enthalpy H. Replacing $\Delta\Phi_j(\eta_j,\delta_j)_{x,y}$ with $\Delta H_j(\eta_j,\delta_j)_{p,S}$ and dividing obtained expression by RT, we get a similar to (21) map

$$\Delta h_j(\eta_j,\delta_j) - \chi_j\varphi(\delta_j,\pi) = 0. \qquad (26)$$

Here $\Delta h_j(\eta_j,\delta_j)=\Delta H_j(\eta_j,\delta_j)_{p,S}/RT_{ad}$, $\chi_j=\alpha_j/RT_a$ is the adiabatic factor analogous to the isothermobaric factor $\tau_j$, and $T_{ad}$ is adiabatic thermodynamic temperature. So far we had no chance to proceed further in this direction. In the most of real problems the adiabatic processes are highly energetic, and adiabatic map perhaps might not make a great difference comparing to classical equations. Further on in this paper we will investigate exclusively the features of isothermobaric maps and their applications.

**The Chemical System Domain of States**

The growth factor plays critical role in dynamical systems being in charge of their evolution. Graphical dependences between $\delta_j$ and growth factor $\tau_j$ are known as bifurcation diagrams. A continuous set of such diagrams, that in our case may be called the *chemical system diagrams of states*, represents *chemical system domain of states*. For example, a set of selected isothermobaric bifurcation diagrams for the system with reaction (9), whose states are defined by (21) is shown in Fig.2. If initial reactant ratio and thermodynamic parameters are varied, the diagrams sweep the whole domain space.

The bio-populations growth theory, where the basic parents/kids and birth/death relations are irreversible, suggests $0\leq\Delta_j\leq 1$. Chemical equilibrium shifts are 2-directional: population of reactants may give up to the products and vise versa, and in some cases it makes sense to admit $\delta_j<0$, a shift towards the logistically exhausted reacting mixture. Two-way bifurcation diagram with the shifts of both signs is shown in Fig.3. Fat lines on the graphs are either to underline a selected at random one of several curves like in Figs.2,3, or just result from larger density of points on the curves. The diagrams have several distinctive areas that have specific meanings for the chemical systems. Three of them can be seen in Fig.2. The area of zero deviation from TdE, where TdE is a strong point attractor with the locus on abscissa, arrives first. The second is the area of the open equilibrium (OpE) where $\delta_j\neq 0$ but the map still has only one solution. The domain curves in both areas are the locuci of particular solutions to map (21) where the iterations converge to fixed points, and after sufficient iterations $\delta_{j(n+1)}-\delta_{jn}<\varepsilon$. When thermodynamic branch becomes unstable and the second solution arrives, the bifurcations area with multiple values of $\delta_j\neq 0$ and multiple states comes into existence. Further increase of $\tau_j$ smoothly leads to chaos via a series of bifurcation points with doubling bifurcation branches.

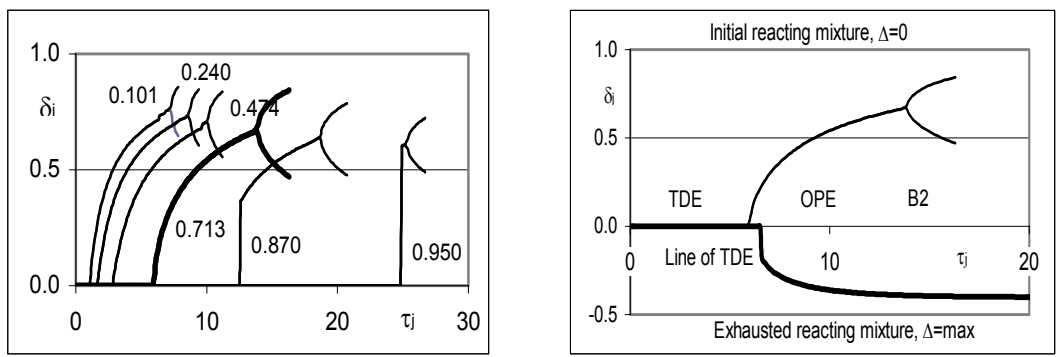

Fig.2. (L) Chemical system domain of states, reaction (9). The numbers are the $\eta_j$ values.
Fig.3. (R) Two-way bifurcation diagram, reaction (9) at 373.15 K, $\eta_j=0.713$.

The OpE area represents a watershed between the point attractor, whose basin is probabilistic kingdom of classical chemical thermodynamics at TdE, on one side, and the strange attractor, the "far-from-equilibrium" wild republic of states-by-chance beyond the OpE limit. The least expected and very interesting result of the new theory is that the *TdE area in open systems is a stretched far enough piece of abscissa with $\delta_j=0$, contracting to a point in isolated systems*. The states along this area are not sensitive to increase of $\tau_j$, resembling so called indifferent equilibria [18].

**The Area Limits**

The area limits $\tau_{TdELim}$ and $\tau_{OpELim}$ unambiguously depend on $\Delta G_j^0$ (and on $\eta_j$) (Fig.4). In systems with robust reactions (very negative $\Delta G_j^0$, $\eta_j \to 1$) the most typical are the TdE and, if achievable, open equilibrium areas, for non-robust reactions (slightly negative or positive $\Delta G_j^0$, $\eta_j \to 0$, e.g., organic and biochemical systems) the bifurcations area may be important as well. Both shown in Fig.4 curves asymptotically meet somewhere at very negative $\Delta G_j^0$, where the OpE area degenerates and the limit value of $\tau$ is so large that one can barely hope to overcome it, and also at a certain point on the positive side of $\Delta G_j^0$. Similar pictures with closed loops, formed by the TdE and OpE limit curves, were found for reactions of various stoichiometry. The TdE area endpoint, or

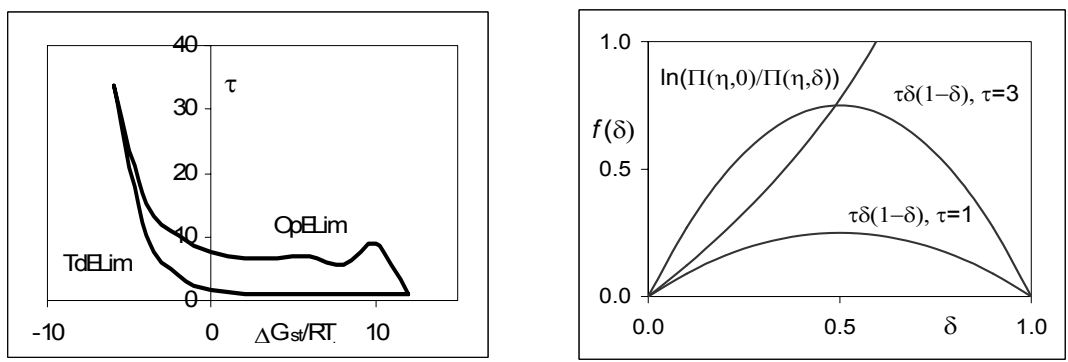

Fig.4 (L). The area limits on the diagram of state, reaction (9).
Fig.5 (R). Terms of map (21), reaction (9), $\eta_j=0.87$ (T=348.15K).

*classical* limit is perhaps the most important. At this point the system starts to deviate from TdE and responds to the external thermodynamic force by changing its state from isolated to open. It is the real border between the classical and the new theories. The limit value can be found out either iteratively or analytically. Indeed, as it can be seen in Fig.5, both terms of map (21) have at least one joint point at $\delta_j=0$ (Fig.6), i.e. at the zero point of coordinates, providing for a trivial solution and retaining the system within the TdE area. The curves may cross somewhere else at least one time more; in this case the solution will differ from zero and number of the roots will be two. There is no intersection if

(27) $\qquad d(\tau\delta\Delta)/d\delta < d\{\ln[\Pi_j(\eta_j,0)/\Pi_j(\eta_j,\delta_j)]\}/d\delta.$

This condition of TdE leads to a quite universal formula for the TdE limit as

(28) $\qquad \tau_{TdELim}=1+\eta_j\Sigma\,[\nu_{kj}/(n^0_{kj}-\nu_{kj}\eta_j)],$

where $n^0_{kj}$ – initial amount and $\nu_{kj}$ – stoichiometric coefficient of k-participant in j-subsystem. For full derivation of (28) see [19].

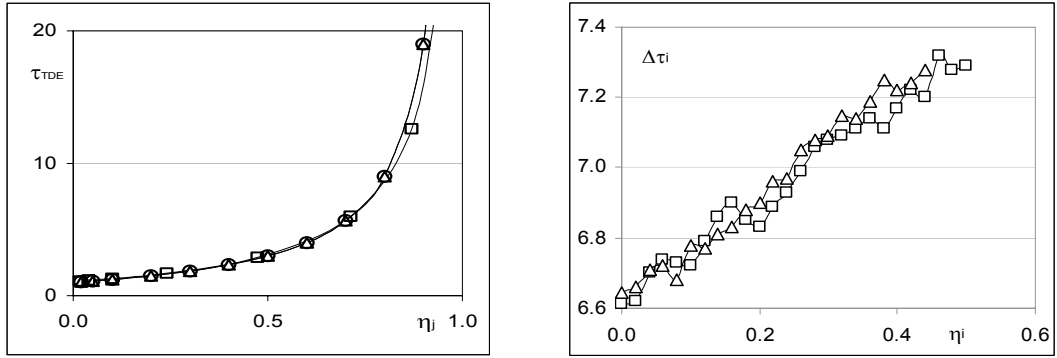

Fig.6. (L) Calculated (o, △) by equations (28) and (29) and simulated (□)values of $\tau_{TdELim}$ vs. $\eta_j$ for reaction (9), $\delta_j>0$.

Fig.7. (R) Extent of the OpE area as $\Delta\tau_i=(\tau_{iOpELim} - \tau_{iTdELim})$, □ and △ correspond to reaction aA+bB=cC, with the sets of stoichiometric coefficients {1,1,1} and {1,2,1} respectively. Thermochemical data are numerically the same as for reaction (9).

Though the area with $\delta_j<0$ is more complicated, in many cases formula (28) is still valid. In a particular case, if initial participant amounts in reaction A+B=C are {1,1,0} moles, expression (28) simplifies to

(29) $\qquad \tau_{TdE}=(1+\eta_j)/(1-\eta_j).$

Fig.6 shows the $\tau_{TdELim}$ values for reaction (9) obtained by iterative process and calculated by formulae (28) and (29), depending on $\eta_j$. Other options to find $\tau_{TdELim}$ can be found in [20].

In computer experiments within a wide range of reaction stoichiometry and initial compositions, two observations regarding the classical limit were done:

- *when $\eta_j\to 0$, the classical limit tends to the sum of stoichiometric coefficients of the reaction products:* $\tau_{TdELim|\eta_j\to 0}\to\Sigma\nu_{pj}.$
- *regardless the $\Sigma\nu_{pj}$ value, the larger is $\eta_j$ the farther is the classical limit from the zero point of abscissa; at $\eta_j\to 1$ the classical limit tends to infinity:* $\tau_{TdELim|\eta_j\to 1}\to\infty.$ *This is the case of absolute irreversibility.*

A clear trend $\tau_{TdELim}\to\infty$ as $\eta_j\to 1$ also follows from (29) and meets the simulation data. The first observation is relevant to the systems with thermodynamically not-robust reactions, the second - to the systems with robust reactions. For the system with reaction

(30) $\qquad A+B=\nu_cC+\nu_dD$

the first observation is illustrated in Fig.8, $\Sigma\nu_{pj}$ was varied from 1 to 8. The $\Sigma\nu_{pj}$ value also affects the diagram shape and location within the system domain of states (Fig.9).

The second observation means that the more robust is the reaction, the more powerful should be the external impact to move the system out the classical area. *This observation turns the thermodynamic equivalent of transformation $\eta_j$ on an effective criterion of chemical irreversibility.*

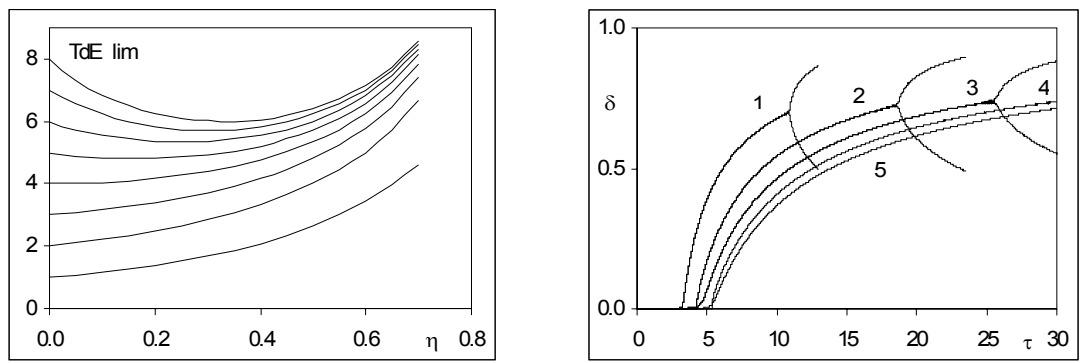

Fig.8. (L) TdE area limit vs. $\eta_j$, reaction (30), $\Sigma\nu_{pj}$ for each curve equals to the TdE$_{lim}$ at its crossing point with ordinate.

Fig.9. (R) Bifurcation diagrams for various $\Sigma\nu_{pj}$, reaction (30), $\eta_j$=0.573, numbers at the curves are the $\Sigma\nu_{pj}$ values.

The OpE limit physically means the end of the thermodynamic branch stability under the stress, caused by external impact. At this point the Liapunov exponent changes its value from negative to positive, and the iterations start to diverge [4]. If map (21) is written in the form of

(31) $$\delta_{j(n+1)}=f(\delta_{jn}),$$

the OpE limit can be found as a point along the $\tau_j$ axis where the $|f`(\delta_{jn})|$ value changes from $(-1)$ to $(+1)$. So far we do not have ready formula for this limit, but it can be found by iterative calculation of the $\delta_j$–$\tau_j$ curve at $\tau_j > \tau_{TdE}$. As it was found for two different sets of stoichiometric coefficients (shown on the capture), dependence of the OpE area extent (i.e. difference $\tau_{OpELim} - \tau_{TdELim}$) on the $\eta_j$ seems to be close to linear, though with essential deviation, (Fig.7). It is unclear how general is that statement, but, obviously, the larger is $\eta_j$ the stronger is the system resistance against bifurcations. It is clear that the OpE limit is *the threshold of the Zel'dovich theorem applicability to open chemical systems*.

Domain of states as a whole is restricted along positive $\delta$–axis by unity – no one chemical reaction cannot be pushed back below its initial state; for $\delta_j<0$ its lower limit is defined by the reactants ratio and actually is the reaction logistic limit.

**Case of Complicated Chemical Systems**

As it was mentioned above, the LCR complicates with the system intricacy. Rewriting expression (21) as

(32) $$\ln[\Pi_j(\eta_j,0)/\Pi_j(\eta_j,\delta_j)] - \tau_j\varphi(\delta_j,\pi) = 0.$$

we obtain a general condition of isothermobaric chemical equilibrium. Now a new feature of the basic map is that the shift function $\varphi(\delta_j,\pi)$ and solutions to (32) essentially depend on $w_0$: this coefficient splits solutions by 2 groups, seemingly relevant to quite different types of chemical systems – *weak*, $w_0=0$, and *strong*, $w_0\neq 0$. Using mathematical induction, at finite $\pi$ and $w_0\neq 0$ one can easily obtain

(33) $$\varphi_u(\delta_j,\pi)=\delta_j(1-\delta_j^\pi),$$

map (32) turns to

(34) $$\ln[\Pi_j(\eta_j,0)/\Pi_j(\eta_j,\delta_j)] - \tau_j\delta_j(1-\delta_j^\pi)=0.$$

At $\pi=1$ map (33) elegantly returns back to (21). Applying the same to the weak systems with $w_0=0$, we get

(35) $$\varphi_z(\delta_j,\pi)=(1-\delta_j^{\pi+1}),$$

and now the weak system map is

(36) $$\ln[\Pi_j(\eta_j,0)/\Pi_j(\eta_j,\delta_j)]-\tau_j(1-\delta_j^{\pi+1})=0.$$

The weak systems are prone to bifurcations even if $\pi=0$.

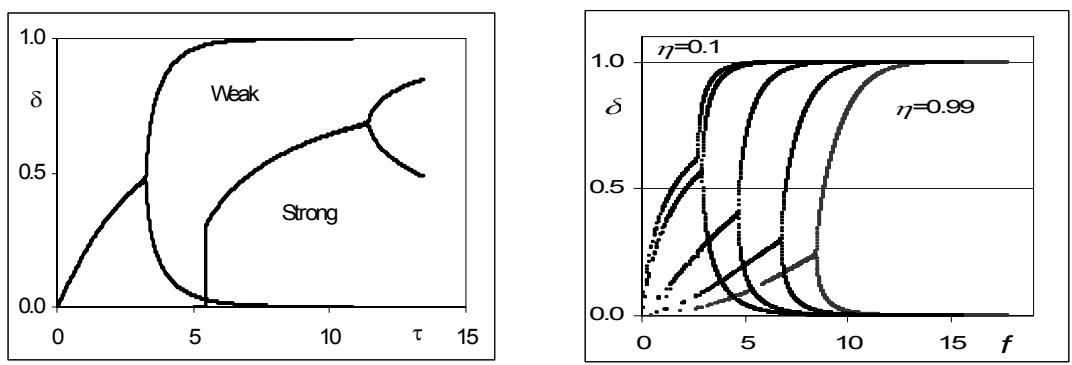

Fig.10. (L) Weak and strong systems bifurcation diagrams, reaction (9), $\eta=0.8$, $\pi=1$.
Fig.11. (R) Dynamic domain of states, reaction (9), weak, $\pi=1$.

Fig.10 gives qualitative comparison of appropriate bifurcation diagrams. The strong system shows essential capability to resist external impact along the TdE area, featuring totally classical behavior below the TdE limit. The weak systems skip the TdE area and do not behave themselves classically except the $\tau_j=0$ point. Bifurcation branches in the strong system diverge slowly, while in the weak system the external force is pushing them to the limit values of $\delta=0$ and $\delta=1$ almost vertically right away from bifurcation point. The weak system resembles a trigger. The said is also well illustrated by Fig.14 and Fig.15. One should notice that now the right hand side of expression (25) for equilibrium constant splits by $\Pi_j(\eta_j,\delta_j)\exp[\tau_j\delta_j(1-\delta_j^{\pi})]$ for the strong and $\Pi_j(\eta_j,\delta_j)\exp[\tau_j(1-\delta_j^{\pi+1})]$ for the weak chemical systems.

At the moment we are not able to tell *a priori* between the specific chemical reactions with the weak or the strong solutions. At a glance, one might turn the strong system to its weak counterpart by mere pushing the TdE limit to zero. Unfortunately, Fig.4 (as well as Fig.6) shows that the TdE limit in the strong systems tends not to zero but to a typical for the traditional logistic map solutions finite value. Most probably, the weak and the strong types of chemical systems may be smoothly transformed each to other passing through intermediate states by variation of the $w_0$ value.

We urge the reader again to distinguish between the reaction robustness, defined by $\Delta G^0_j$ or $\eta_j$, and the system strength, defined by the type of system reaction to the external impact and therefore by the type of its diagram of state. Further on in this paper we will deal mainly with the strong chemical systems if the opposite isn't stated.

**Dynamic Bifurcation Diagrams**

Traditional bifurcation diagrams give a good idea of what happens to the system with increase of the control parameter $\tau_j$. However, such a description in case of chemical systems contains not enough clear physico-chemical information. Trying to relate the system's response to its reason, we have developed a new type of bifurcation diagrams - shift–force, or *dynamic* bifurcation diagrams (d-diagrams); $\delta-\tau$ diagrams are *static* (s-diagrams). As it follows from (15), internal TdF via change of Gibbs' free energy in finite differences is $F_{ij}=-\Delta G_j/\Delta_j$. Dividing maps (34) and (36) by $\Delta_j$, after simple algebraic transformations of the second terms one can get *dynamic state diagrams*

(37) $\quad\{\ln[\Pi_j(\eta_j,0)/\Pi_j(\eta_j,\delta_j)]\}/(1-\delta_j) - \tau_j\delta_j\Sigma_{0-\pi}\delta_j^p = 0,$

for the strong, and

(38) $\quad\{\ln[\Pi_j(\eta_j,0)/\Pi_j(\eta_j,\delta_j)]\}/(1-\delta_j) - \tau_j\Sigma_{0-(\pi+1)]}\delta_j^p = 0$

for the weak systems. If $\pi$ is large, the second terms in both cases tend to expressions $\tau_j\delta_j/(1-\delta_j)$ and $\tau_j/(1-\delta_j)$ respectively with the same expressions for the bound affinity as the first terms and reduced by RT external TdF for the second terms.

Maps (37) and (38) express condition of chemical equilibrium for two types of chemical systems as explicit balance of thermodynamic forces. Their graphical solutions constitute the *dynamic domains of states*, partly shown for a weak system in Fig.11. D–diagrams skip the TdE area ($\delta_j=0$ turns the TdF to zero), the OpE area stems out immediately from zero point of coordinates. They are only slightly different for the weak and the strong systems, that difference is less expressed than the difference between appropriate static bifurcation diagrams. Though the shape of d–diagram resembles s–diagrams of the weak systems, one should keep in mind that they are plotted against different values. That type of maps and diagrams are mostly important for various applications and thermodynamic simulation of complex equilibria.

**Discrete vs. Classical Shift-Force Curves**

It is possible to create shift-force diagram exclusively within the classical theory, imitating shifts and TdF. To do so, one should choose an arbitrary equilibrium state of isolated system and accept is as the basic TdE with $\delta_j=0$ and certain $\eta_j$ value, and then vary $\delta_j$ and calculate corresponding TdF as the first term of the left hand side of (37), holding the chosen $\eta_j$ unchanged. That will create a classical imitation of the shift–force dependency. Comparison with a d–diagram, is given in Fig.12. The curves are pretty close, the DTD diagram heads into bifurcations area and the fake doesn't. In the positive shift quadrants DTD restricts the system one-valued shift from TdE by the bifurcation point, while classical theory greenlights full return to the initial reaction state. Fig.13 presents data for reactions of various double oxides with sulfur, based on standard changes of Gibbs' free energy and conventional thermodynamic simulation; both are totally classical and considered reliable. Calculation and plotting the data on Fig.13 followed the above described procedure for double-oxides, numerical results are placed in Table I. Zero point was taken logically for an abstract oxide binding itself, i.e. MeO·MeO. Every point in Fig.13 may be considered just another value of external TdF and corresponding shift from TdE, regardless the chemical elements comprising these double oxides. Interestingly enough that the points are gathering along imaginary thermodynamic branch in the OpE area but no bifurcations can be found. The first part of the correlative curve, below $\delta_j \approx 0.65$ in Fig.13 may be interpolated as linear with a moderate error, in compliance with the data in Fig.1.

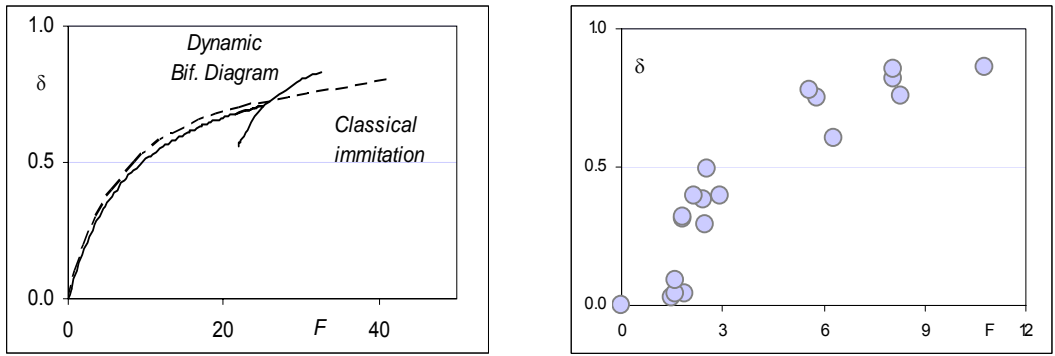

Fig.12. (L) D–diagram and its classical imitation, reaction (9), 398.15K, $\eta_j=0.474$.
Fig.13. (R) Shift-force classical imitation: reactions of various double oxides with sulfur.

Table I. The shift-force classical imitation for various double oxides.

|  | $\Delta G_j^0/(\Delta_j\cdot RT)$ | $\delta_j$ |  | $\Delta G_j^0/(\Delta_j\cdot RT)$ | $\delta_j$ |
|---|---|---|---|---|---|
| MeO·MeO | 0.00 | 0.00 |  |  |  |
| FeO·SiO2 | 1.50 | 0.03 | CoO·Cr2O4 | 2.94 | 0.39 |
| FeO·Cr2O3 | 1.86 | 0.04 | NiO·WO3 | 2.55 | 0.49 |
| NiO·TiO2 | 1.58 | 0.05 | FeO·Al2O3 | 6.28 | 0.61 |
| NiO·SiO2 | 1.60 | 0.09 | MnO·SiO2 | 5.81 | 0.75 |
| NiO·Cr2O3 | 2.51 | 0.29 | PbO·B2O3 | 8.28 | 0.76 |
| CoO·Fe2O3 | 1.82 | 0.31 | MnO·Fe2O3 | 5.58 | 0.78 |
| FeO·Fe2O3 | 1.84 | 0.32 | PbO·SiO2 | 8.07 | 0.82 |
| CoO·TiO2 | 2.44 | 0.38 | PbO·WO3 | 8.07 | 0.85 |
| CoO·WO3 | 2.14 | 0.39 | PbO·TiO2 | 10.81 | 0.86 |

**Emergence of Bifurcations in Chemical Systems**

We equalize all the weights $w_i$ in the LCR (10) to unity beginning with p=1 just because we don't know their values yet. What changes if we vary the values? Results in this chapter were obtained for the simplest reaction $A^*=A+\Delta E$, like photoreactions or phase transitions, varying only $w_1$ in

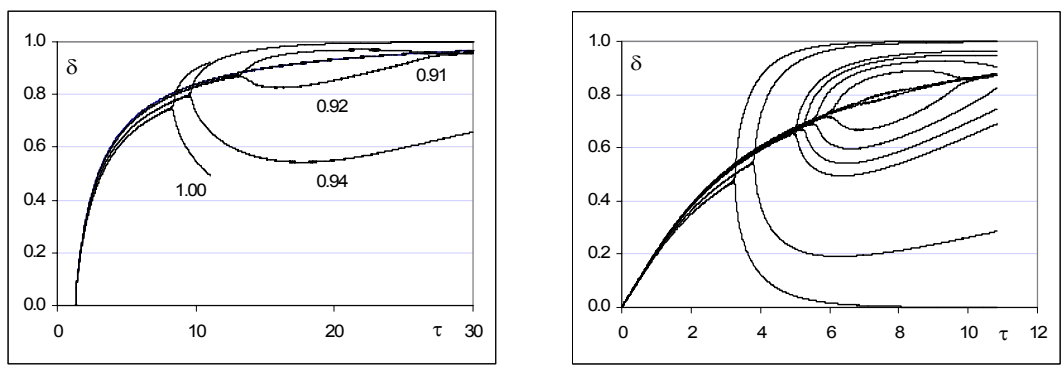

Emergence of bifurcations on the static diagrams, varying $w_1$ in the LCR (32):
Fig.14. (L) Strong system, $w_1$ values are shown at the curves.
Fig.15. (R) Weak system, $w_1$= 0.75, 0.76, 0.77, 0.78, 0.79, 0.80, 0.9, 1.0, inner to outer curves.

$\varphi(\delta_j,\pi)$ (map (32)). It allowed us to observe emergence and development of pitchfork bifurcations on s– and d–diagrams as well. Fig.14 and Fig.15 show the birth and evolution of bifurcations in respectively strong and weak systems, $\eta=0.8$, $\pi=1$, varying $w_1$. The similarity of those processes in both pictures is remarkable – bifurcations start at a certain value of $w_1$ with a loop bifurcation, resembling a bud on thermodynamic branch at the OpE area endpoint. They open up slowly like a contour of a flower whose sidelines then gradually evolve into typical for each type of a system pitchfork bifurcation branches. In similar cases the strong systems show larger shift from TdE.

**Complexity Parameter and Diversity of States in Chemical Systems**

Though it cannot be called evolution in direct sense, our results show that under certain circumstances chemical systems can evolve in a way similar to bio-societies, following well studied pattern. Of course this evolution is changing the states of the chemical system, and after all the system will comprise nothing else but chemical species.

The prerequisite for such an evolution is the system openness, and the external TdF should be mighty enough to push the system beyond the OpE limit. Earlier shown pictures abounded with examples, relevant to simple systems. A picture for static diagrams with varying complexity is presented on Fig.16. As $\pi$ increases, the diagram shape and $\tau_{TdELim}$ stay unchanged; one can see bifurcation points, climbing up the vertical stem, and the shrinking OpE area. Typical d–diagram of the evolving chemical system includes a sharply raising from zero and asymptotically tending to saturation ($\delta_j \to 1$) unique *evolution trunk*, a locus of bifurcation points. Latter are moving up the trunk as the system complexity increases - the higher is the system complexity, the larger is the corresponding shift and the larger should be the external TdF to achieve the bifurcation point (Fig.17). The trunk continuity from $\pi=1$ to very high values speaks for a self-similarity of the graphical solutions to maps (37) and (38), the trunk curve is a unique solution of these maps at $\pi=1$, that is with the second term equal to $\tau_j\delta_j$.

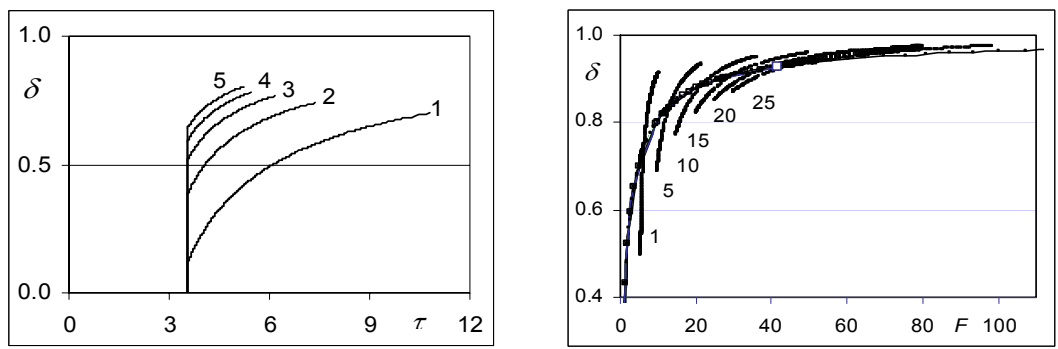

Fig.16. (L) Bunch of static bifurcation diagrams, stemming out from the same TdE limit point into the OpE area, reaction (9), $\eta=0.57$, numbers at the curves - the $\pi$ values.

Fig.17. (R) Dynamic bifurcation diagram, reaction (9), $\eta=0.20$. One can see the trunk line with growing out bifurcation "wings", numbers at the branches are $\pi$-values.

One can see the clockwise rotation of the diagram wings on Fig.17 with increase of $\pi$ and their asymptotic folding towards the evolution trunk, eventually they merge. Does it mean that the complex chemical system evolution ends at the merge point?

Simultaneous variations of complexity and system/reaction parameters lead to further ramifications of bifurcation diagrams, instead of singular points we see clusters of them sitting on the evolution trunks as the reaction products stoichiometric coefficients vary. An example for reaction A+B=cC with varying $\pi$ and c is given in Fig.18. The clusters contain closely sitting bifurcation points for various $\eta$ (0.05,…,0.97), the wings rotate clockwise while the cluster axes rotate counterclockwise. Very probably the upper $\pi$ and F values we used were not realistic, just allowing us to show the wider picture.

**System Complexity and Location of Bifurcation Point Period 2**

Despite of some serious quantitative approaches to the systems self-organization (e.g., [6]), perhaps the only universal conclusion on the chemical system behavior in the "far-from-equilibrium" area is a tautology that sophisticated behavior should be expected in complicated systems.

Le Chatelier Response (10) sets a certain relationship between the chemical system complexity and its response to the external force. Investigating evolution, we have found a very interesting dependence of the shift value at the bifurcation point $\delta_{OpELim}=\delta_\pi$ upon the $\pi$ value (Fig.21). The linearity of $\delta_\pi$ vs. $\ln\pi$ is obvious - we have encountered a new empirical rule, to the best of our knowledge never observed before and confirmed by a number of our simulations, that is

(39) $$\delta_\pi = \delta_0 + \beta \ln\pi.$$

Expression (39) is to replace the linguistic variable "far-from-equilibrium" with a precise number: *deviation of the chemical system from TdE where its thermodynamic branch becomes unstable is directly proportional to logarithm of the system complexity parameter*.

Term $\delta_0$ corresponds to $\pi=1$ that we have analyzed before; its value slightly depends upon $\eta$, but eventually, as $\pi$ increases, all lines converge. It also follows from this rule that *the larger is $\pi$, the less the complex system is prone to changes*. The observation is the same for both types of chemical systems, the strong and the weak, as well as for the static and the dynamic bifurcation diagrams. Thermodynamic force, enough to move the system beyond bifurcation point and to destabilize thermodynamic branch, the *stability threshold of thermodynamic branch vs. external force* equals to $[\tau_\pi\varphi(\delta_\pi,\pi)/(1-\delta_\pi)]$. From the energetic point of view, the force equals to the energy sufficient to push chemical reaction back towards its TdE.

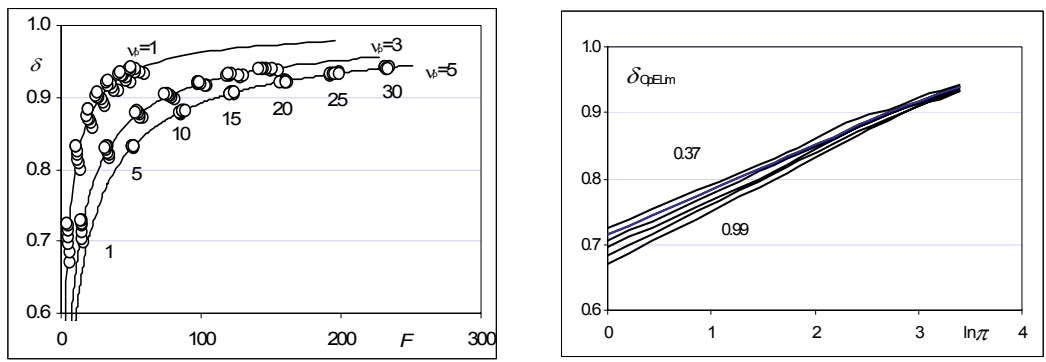

Fig.18. (L) Dynamic evolution trunks with nesting on them clusters of bifurcation points, reaction A+B=cC, varying $\eta$, $\pi$ and c ($\nu_p$). Numbers at the graph show the sequential $\pi$ values.
Fig.19. (R) The shift value at bifurcation point vs. $\ln\pi$, reaction (9), numbers at the slopes are $\eta$.

**Discrete Thermodynamics and Coefficients of Thermodynamic Activity**

The role of thermodynamic activity coefficients in the classical theory was discussed above. If deduced from probabilistic considerations, they could be more organically woven into the classical canvas of thermodynamics. Indeed, if only one reaction runs in a system, the outcome is defined by a probability of the reaction participants to collide, or by the product of their *a priori* probabilities to occur simultaneously at the same point of space. These probabilities equal to their concentrations or mole fractions in the reacting mixture, thus we arrive at the traditional mass action law. The situation gets worse if several coupled chemical reactions are running simultaneously in the system. Let $\{R_{1k},R_{2k},R_{3k}, …\}$ be a set of the only possible events on the reaction space $S_j$, competing for the k-component. Because its amount in the system is restricted, the events outcomes are mutually dependent, and now the mass action law comprises *conditional* probabilities. Let event $A_k$ consists in occurrence of any $R_{jk}$ on the space $S_j$; then the statement known as Bayes' theorem [22]

(40) $\qquad p(R_{ik}|A_k)=p(R_{jk})p(A_k|R_{ik})/\Sigma[p(R_{jk})p(A_k|R_{jk})]$.

defines conditional probability of $R_{ik}$ given $A_k$. A probabilistic activity coefficient is a ratio

(41) $\qquad \gamma^{\nu_{kA}}_{jk} = p(R_{jk}|A_k)/p(R_{jk})$.

This is exactly the coefficient of thermodynamic activity, introduced by Lewis in a different context. Combination of the coefficients, related to components of the same reaction, or to the dwellers of the same subsystem, designed exactly as appropriate mole fractions product, form excessive thermodynamic functions in change of Gibbs' free energy, represented by the third term in following expression

(42) $\qquad \Delta G_j = \Delta G^0_j + RT\ln\Pi_j(x_{kj}^{\nu_{kj}})] + RT\ln\Pi(\gamma_{kj}^{\nu_{kj}})$,

$x_{kj}$ are mole fractions, $\nu_{kj}$ are the stoichiometric coefficients. Though the Bayes' theorem weighs prior information with empirical evidence, it is still the best tool to demonstrate the probabilistic meaning of activity coefficients. Comparison between map (21) and equation (42) with $\Pi_j(\eta_j,\delta_j)=\ln\Pi_j(x_{kj}^{\nu_{kj}})$ prompts us to suggest the second term in (21) and the third term in (42) to carry the same functionalities, both reflect the system openness and external impact. After reducing them by RT

(43) $\qquad\qquad\qquad\qquad\qquad\qquad\qquad\tau_j\delta_j(1-\delta_j)=-\ln\Pi(\gamma^{\nu}{}_{kj})$.

For the simplest case with the only one activity coefficient per subsystem

(44) $\qquad\qquad\qquad\qquad\qquad\qquad\qquad\delta_j=(1/\tau_j)[(-\ln\gamma_{kj})/\Delta_j]$.

To validate expression (44), we carried out simulations for reaction

(45) $\qquad\qquad\qquad\qquad\qquad\mathbf{2CoO+4S=CoS_2+CoS+SO_2}$

at p=0.1 Pa, T=1000K with reactants taken in stoichiometric ratio, varying activity coefficients of CoO. For comparison we have simulated appropriate s–diagrams. The external TdF at equilibrium was counted in two ways – first as $F_{ej}=\ln[\Pi_j(\eta_j,0)/\Pi_j(\eta_j,\delta_j)]/\Delta_j$ and then as $F_\gamma=(-\ln\gamma_{kj})/\Delta_j$. The force-shift dependence for this reaction is shown in Fig.20. The above found similarity and equation (44) form a ground for independent method to find out coefficients of thermodynamic activity for various applications. Though they are out of the DTD concept, using them in some cases is still simpler. From this point of view, what we have done in DTD with regard to activity coefficients may serve as a fresh view at their nature and an alternative method of their calculation. More details regarding the method may be found in [20].

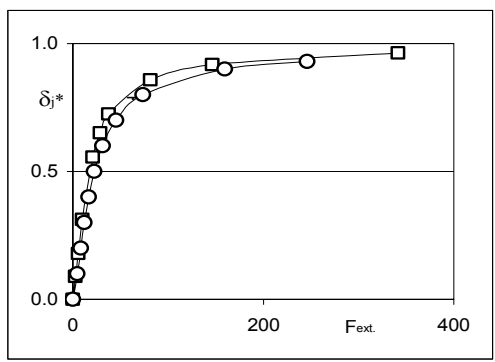

Fig.20. Force-shift graphs, reaction (45), 1000K, curves are relevant to dimensionless TdF as $F_{ej}$ (□) and as $F_\gamma$ (o) respectively.

**Discrete Thermodynamic Simulation**

A habit, born mainly by absence of a descent theory of the open systems and deeply planted in current thermodynamic mentality, prompts researches to equalize chemical equilibrium and TdE. It is simple and convenient − most famous computer software for thermodynamic simulation is based on this approach. But it is not true. That's why the software is able to provide a realistic quantitative analysis only for a few objects, mainly in adiabatic systems and related to reactions with huge negative $\Delta G^0$. Discrete thermodynamics changes the situation, opening opportunities for advanced thermodynamic simulation and analysis of real chemical systems in a wide range of $\Delta G^0$.

Based on map (21), thermodynamic simulation of complex chemical systems takes into account their nature as comprising mutually open entities. State of each entity is fully determined by its shift from TdE. Domain of states for each entity is a totally individual set of graphical solutions to this entity map (21); stoichiometry and thermochemical characteristics of chemical reaction, running within a subsystem, as well as initial composition of the reacting mixture together lead to a certain bifurcation diagram. The subsystem interactions with its compliment define the subsystem's state as a point in its TdE or OpE areas with unilaterally linked together $\tau_{ij}$ and $\delta_{ij}$. Joint solution for the maps of states (21) of individual subsystems at restricted resources gives the coordinates to the all entity states at complex equilibrium.

No surprise that parameter $\tau_j$ took a great deal of attention in this work. If the $\tau_j$ value falls into the TdE area, the subsystem behaves itself classically. It is clear that regarding thermodynamic simulation the OpE area will be the primary object of our interest. To show the numbers and to illustrate how it works in simple cases, we have carried calculations for reactions in mixtures with double oxides along the previously bitten path, using the standard changes of Gibbs' free energy of their formation from single oxides were used to find corresponding $\tau_j$ values. The chain of calculations was as follows: $\Delta G^0_{f(MeO \cdot RO)}$ was considered the excess function for any of the two single oxides; it gave us the $\gamma_j$ value; then $\tau_j\delta_j$ was obtained as $(-\ln\gamma_{kj})/\Delta_j$, and finally the bifurcation diagram was used to get the appropriate $\tau_j$ value. Homological series of reactions

(46) $\qquad\qquad\qquad\qquad\qquad\qquad\qquad\qquad$ **2CoO·RO+4S=CoS$_2$+CoS+SO$_2$+RO**

were simulated at p=0.1 Pa, T=1,000K with reactants taken in stoichiometric ratio, R={TiO$_2$,WO$_3$,Cr$_2$O$_3$}. These oxides do not react with sulfur, and we have two competing for Co subsystems - one with reaction (CoO+RO) and another with (2CoO+4S). Classical thermodynamic simulation was performed on reaction (47) with variable CoO activity coefficients using the HSC simulation program [23]. To compare the results, $(-RT\ln\gamma_j)$ was considered the excess function. Simulation data are shown in Table II. The DTD data are the graphical solutions to corresponding maps (21); their logic is clear from Fig.21. The curves are plotted in coordinates $\Delta_j$ vs. either $\Delta G_j/RT=\ln[\Pi_j(\eta_j,0)/\Pi_j(\eta_j,\delta_j)]$ (ascending curves) or vs. the parabolic term $\tau_j\delta_j(1-\delta_j)$ (descending curve) for double oxides formation reactions from oxides. Intersections of the ascending curves with the descending curve give the numerical values of corresponding reaction coordinates. The CoO ascending curve hits the ordinate at $\Delta_j=1$ (or $\delta_j=0$), marking the undisturbed TdE. The major point to pay attention to is a difference between 2 sets of reaction shifts. In classical simulation all reaction extents are equilibrium; all presented in the table classical "shifts" are artifacts, while the DTD shifts are "natural".

Table II. Equilibrium values of reaction extents and shifts in homological series of reaction (46).

| Reactant | CoO | CoO·TiO$_2$ | CoO·WO$_3$ | CoO·Cr$_2$O$_3$ |
|---|---|---|---|---|
| $(-\Delta G^0_{f(CoO \cdot RO)}/RT)$,kJ/m | 0.00 | 3.77 | 6.17 | 7.2 |
| $\Delta$ simulated, HSC | 1.00 | 0.92 | 0.88 | 0.86 |
| $\Delta$ graphical | 1.00 | 0.9 | 0.82 | 0.77 |
| $\delta$ conditional, HSC | 0.00 | 0.08 | 0.12 | 0.14 |
| $\delta$ graphical, DTD | 0.00 | 0.10 | 0.18 | 0.23 |

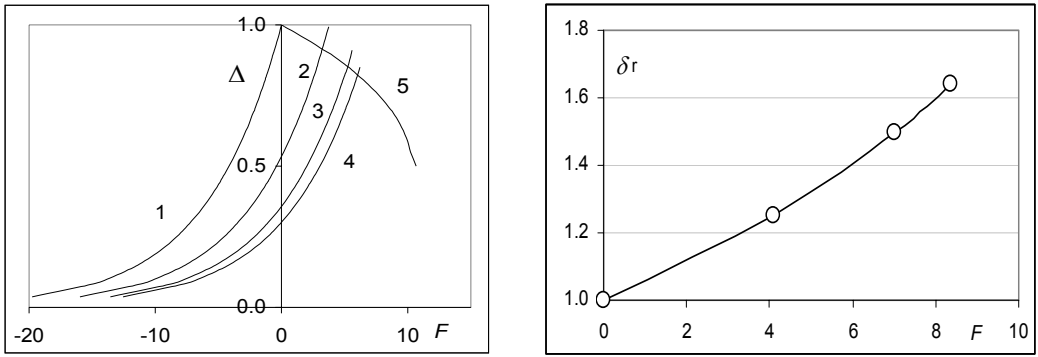

Fig.21 (L). Reaction extent $\Delta_j^*$ vs. TdF. Reaction (46), T=1000 K, ascending curves 1-CoO, 2,3,4-CoO·RO, RO=TiO$_2$, Cr$_2$O$_3$, WO$_3$; descending - the second term of map (21).
Fig.22 (R). Ratio $\delta_r$=(DTD shift/classical shift vs. $F_{ext}$=$(-\Delta G^0_{f(CoO \cdot RO)}/\Delta_j RT)$ (dimensionless).

The most important and quite obvious conclusion is that *equilibrium simulation in discrete thermodynamics allows us in the most cases to avoid usage of thermodynamic activity coefficients.* That extends the advantage of discrete thermodynamic simulation even further because the coefficients are expensive and should be found separately. The ratio between DTD and classical shifts vs. TDF is shown in Fig.22. One can find more details in [19].

**Chemical Oscillations**

Chemical oscillations belong to the most interesting phenomena in current thermodynamics of open chemical systems. The relevant boom in science is well back in time, but it's still one of the most popular corners of science.

We have discovered natural oscillations with variable amplitude within bifurcation branches when external TdF takes on an essential magnitude. Detailed investigation showed that oscillating behavior is typical for the weak case and almost never occurs in the strong model at all. Fig.23 and Fig.24 show the simulation results for the weak system model with reaction (9); the curves are plotted in colors to discern details and to imitate the color changes. Beyond a certain value of TdF

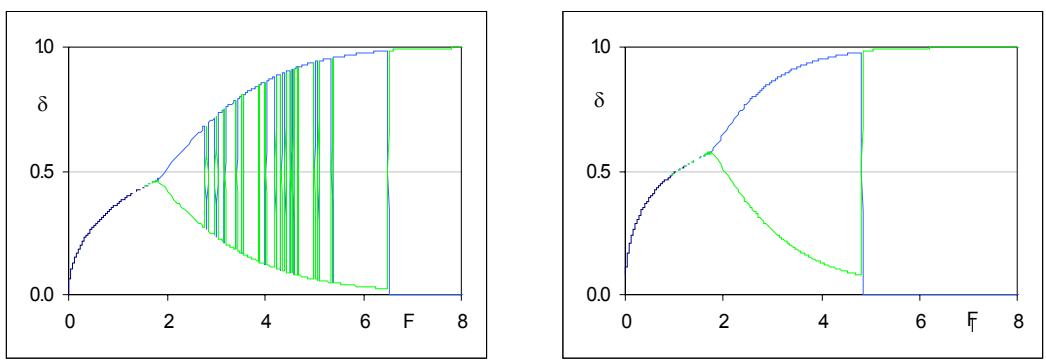

Chemical oscillations, weak system, reaction (9):
Fig.23. (L) T=448.15K, η=0.756.    Fig.24. (R) T=498.15K, η=0.381.

the system experiences oscillations with the force increase. As it follows from our data, DTD chemical oscillations exist within a restricted range of external forces – from η=1 to a point between η=0.756 and η=0.381, where the oscillation picture changes to contain only one switch point (like shown in Fig.24, pay attention to the colors) and then stays the same with variations of the point location along abscissa. The restricted range is in agreement with experimental data for oscillating reactions [17]. Increase of η leads to more ostensible oscillations with variations of the oscillations zone border points within the bifurcation area: the zone begins and ends at larger force values, the pulse durations sharply increase leading to less total pulses amount in the zone, and forcing the color changes to be more pronounced. Fine structure of the DTD oscillations spectrum is shown in Fig.25; the points on the curves correspond to the iterative points in computer simulation. The oscillations lead to short-time conversions of the subsystem's states, accompanied by changes in color; beyond the oscillations zone the states convert. Due to decrease of TdF or in a free fall if TdF turns to zero, the system tends to relax towards the OpE via oscillations that follow the same pattern. Continuous oscillations are typical for non-equilibrium stationary systems with input/output material flows, and are not considered in this paper.

The above found oscillations are relevant to the closed systems with independent sources of external impact. To analyze the systems with auto-oscillations like the Bray-Liebhafsky [24,25] or well-known BZ [26] reactions, one has to consider simultaneously all parts of the chemical system as open systems as it was explained earlier. Observed in this work chemical oscillations follow naturally from DTD, and are predictable thermodynamically without a trace of kinetics or any autocatalytic effects.

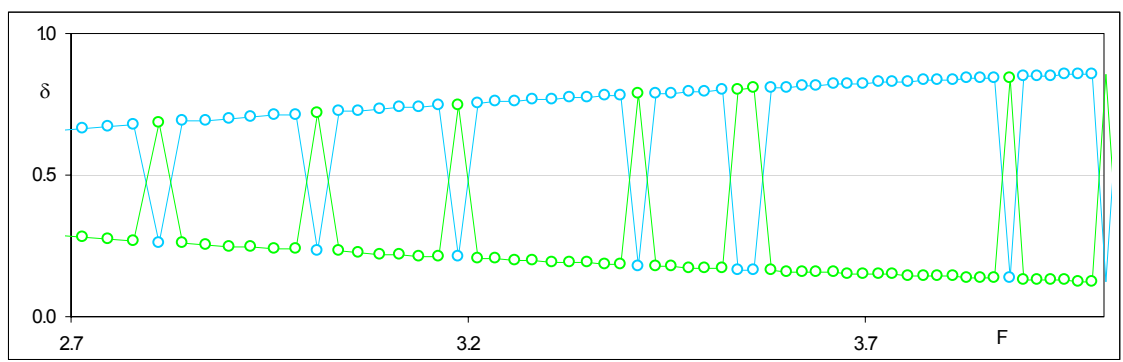

Fig.25. Initial fragment of the bifurcation diagram with chemical oscillations, reaction (9), $\eta=0.756$, horizontal expansion of the graph from Fig.23.

**Quasi-Chemical DTD Application with Oscillations: Discrete Thermodynamics of Lasers**

More or less successful attempts to apply thermodynamics to non-thermodynamic systems were described in several publications. For example, the authors of [27] used modified mass action law for early computer simulation of ecological systems with bio-populations. Discrete thermodynamics may be successfully applied to quasi-chemical systems, where transformation like aA+bB=cC+ΔE takes place, the symbols are not necessarily chemical substances and stoichiometric coefficients. Such a reaction may or may not follow the chemical laws but may obey discrete thermodynamics.

A good example of a weak quasi-chemical thermodynamic system, transforming the input energy into the output, is laser – the system is easily excited and achieves TdE at zero point of the reference frame. As strange as it is, the only found really thermodynamic description of the laser action was famous Einstein's prohibition on the 2-level laser [28], based on Boltzmann statistics and on the equal pump up – drop down probabilities for the excitable atoms. Einstein considered a light emitting process in isolated system with no pumping energy from outside; needless to say that a real lasing device is an open system and lives different life. The ideal laser system should allow to pump up as much as possible amount of atoms from their ground state (subsystem A) to the upper level (subsystem A*), holding them over there as long as necessary until the excited atoms emit light and get back to the ground state.

Let the laser contain 1 mole of the excitable atoms. In absence of the pumping force, the laser stays in TdE between the subsystems. Their TdE populations are $\eta$ (<1) moles of the A atoms and $(1-\eta)$ moles of the A* atoms; the population ratio obeys Boltzmann distribution

(47) $\qquad (1-\eta)/\eta = \exp(-h\nu/kT)$.

In open laser system the upper level population increases due to the energy pumping and absorption

(48) $\qquad A + \varepsilon_{in} \to A^*$,

while stimulated or spontaneous discharge of the accumulated energy via light emission

(49) $\qquad A^* \to A + h\nu$

returns the excited laser population back to the ground.

The heating of laser due to energy dissipation may be prevented by cooling; lasers work at p,T=const, and $\Delta G$ for the A→A* transition is

(50) $\qquad \Delta G = \Delta G^0 + RT\ln[(1-\eta)/\eta]$,

In absence of the pumping force $\Delta G=0$, and

(51) $\qquad (1-\eta)/\eta = \exp(-\Delta G^0/RT)$.

Multiplying both tiers of the exponent power in (47) by the Avogadro number $N_A$ one can get expression for the standard change of Gibbs' free energy value, reduced by RT

(52) $$\Delta g^0 = h\nu N_A/RT,$$
or after gathering the universal constants into $\lambda_\nu = hN_A/R$
(53) $$\Delta g^0 = \lambda_\nu \nu/T,$$
$\lambda_\nu = 47.99$ if the frequency $\nu$ is in terahertz. The amount of the ground level atoms at TdE is, moles
(54) $$\eta = 1/[1+\exp(-\lambda_\nu \nu/T)].$$
For the most important part of the light spectrum, from far infrared to short wave ultraviolet, the $\Delta g^0$ values vary from $-63.040$ for to $-246.720$ dimensionless units at T=300K.

It is well known that in real devices huge majority of the excitable dwellers of any laser are sitting on the ground at TdE. Closeness of $\eta$ to unity makes all frequencies within the practical IR-UV range almost degenerated thermodynamically. The pumping force drives the laser system out of TdE. If x moles of atoms are off the ground level up, the laser proximity to TdE is characterized by dimensionless coordinate of the ground subsystem state

(55) $$\Delta = (\eta - x)/\eta.$$
Obviously at TdE x=0 and $\Delta = 1$. The shift is
(56) $$\delta = x/\eta.$$
The basic map of the laser as a discrete thermodynamic system is expression (38). To include the laser parameters in it, let's neglect the A*−population at TdE ($\eta \approx 1$); then, if x moles of A moved to A*, the population ratio is $\rho = x/(\eta - x)$. Substituting $\eta = x/\delta$ from (56) we get finally $\rho = \delta/(1-\delta)$, obviously equal to $\Pi_j(\eta_j, \delta_j)$ because sum of moles equals to unity. Now we get the dynamic map of the laser states

(57) $$\{\lambda_\nu \nu/T + \ln[\delta/(1-\delta)]\}/(1-\delta) - \tau\Sigma_{0-(\pi+1)]}\delta^p = 0.$$
The third term in (57) is external TdF, in this case the pumping force; after substitution one obtains the shift vs. pumping force relation for lasers.

The 2-level laser with one lasing level corresponds to a system with $\pi = 1$. We have simulated open equilibria of such a laser, the results are graphically interpreted in Fig.26. A typical for the weak system dynamic diagram is well pronounced in $\delta-P_j$ coordinates. As the pumping force and the activated atoms amount increase, the system shifts from TdE along the thermodynamic branch. The branch looses stability at a certain value of the force, splits by two bifurcation branches, and we encounter bi-stability [29] and then oscillations, similar to described in the previous chapter. Originated from alternating up and down jumps, the vertical lines mean instability of bifurcation branches, while the downward lines on Fig.26 correspond to spontaneous transitions and spontaneous emission. All the transition spectra are linear within the studied light frequency range.

The ascending jumps are driven by external pumping power, bringing a part of the laser population to the upper level. Being pumped up, the excited atoms then coherently discharge by spontaneous or assisted irradiation of light along the descending line. In 2-level laser the discharge starts at $\delta \approx 0.6$, this point is far from population inversion. This is an independent proof of the Einstein's prohibition on the 2-level laser, extended to open laser systems and based exclusively on discrete thermodynamics.

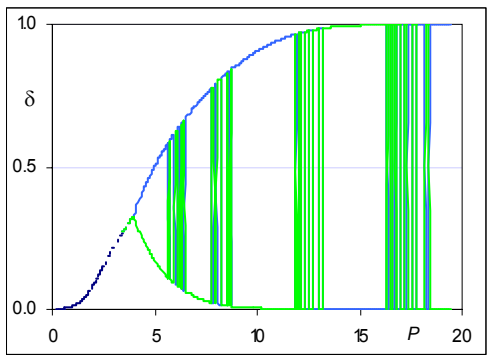  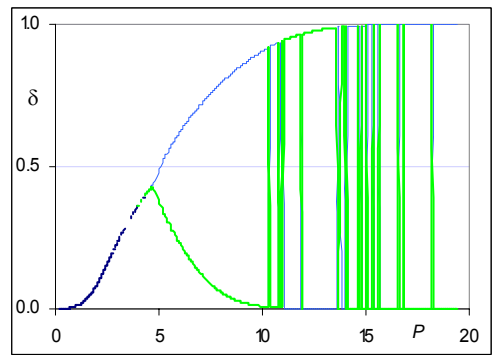

Fig.26. 2-level laser.    Fig.27. 3-level laser.

Simulation results for the 3-level laser with π=2 are plotted in Fig.27. Start of the oscillations zone, or spontaneous emission is pushed essentially to the right on the bifurcation diagram: now it happens at δ≈0.9. This value is in a good match with the 3-level laser practical results. More details and different approach to the 3-level laser DTD can be found in [30].

**Conclusions**

Besides the Le Chatelier response, whose concept originated from quite usual in physics and was modified in this work to account for shift-force relationship in chemical systems, everything else in this work was obtained directly from the well recognized rules and concepts of contemporary thermodynamics. The basic map of states has been derived, not postulated, like, for instance, Schrődinger's equation in quantum mechanics.

The second term in (21), or the exponential factor in (25) make the major difference between discrete and classical thermodynamics of chemical equilibria. This, small at a glance difference is of a critical importance, leading to large differences in chemical systems behavior as the deviation from TdE increases as it is typical for dynamic systems, this term.

DTD clearly presents chemical equilibrium as a system phenomenon, based on the balance of internal and external thermodynamic forces. In general, chemical equilibrium is no less thermodynamic than but not identical to true thermodynamic equilibrium, relevant to isolated systems. We have obtained the basic map (18) and its versions without original intentions to arrive at the logistic map at all, just operating with finite differences that are not strangers for the delta-thinking and delta-speaking thermodynamics. Among others, thermodynamic affinity in finite differences is one of the most important expressions for the theory; equality between it as the partial derivative and as the value in finite differences is critical. For the reactions of species formation from elements, which are the major entities of complex systems thermodynamic analysis, one can easily prove that

(58) $$A_j = -\partial G_j/\partial \xi_j = -\Delta G_j/\Delta \xi_j.$$

Hess' law of constant heat summation [31] allows to extent this result to any chemical reaction.

DTD is intended to extend horizons of understanding. It also offers new methods that make it more capable than classical thermodynamics in solutions to practical problems, relevant to open systems.

It is worthy to mention that some basic ideas of discrete thermodynamics were inspired by and are closely related to the principles of mechanics. Indeed, the DTD concept of binary equilibria repeats famous description of the arch by Leonardo da Vinci - "*The arch is nothing else than a force originated by two weaknesses, ... as the arch is a composite force, it remains in equilibrium because the thrust is equal from both sides.*" [32]. The d'Alembert principle, that turns dynamic problems into static ones due to composing a full set of mutually balanced forces, has led the author to interpret chemical equilibrium in a similar way. The idea of where to look for the equilibrium point in complex systems as sets of open entities was prompted by Gauss principle of minimal constraints, adopted long ago by chemistry in form of the Le Chatelier principle.

So far we have applied DTD mainly to general chemical systems (excluding lasers). Though the DTD basics are solidly formulated, we consider what was done a prelude and expect many interesting and useful DTD applications to electrochemical, bio-populations, and other systems.